\documentclass{webofc}
\usepackage[varg]{txfonts}   
\usepackage[left]{lineno}

\usepackage[frozencache=true]{minted}
\setminted{fontsize=\footnotesize,frame=lines}

\begin{document}
%
\title{Exploring polyglot software frameworks in ALICE with \texttt{FairMQ} and \texttt{fer}}
%
%

\author{
        \firstname{S\'ebastien} \lastname{Binet}\inst{1}\fnsep\thanks{\email{binet@cern.ch}} 
		on behalf of the ALICE collaboration
}

\institute{
           CNRS/IN2P3, LPC, F-63000 Clermont-Ferrand, France
          }

\abstract{%
In order to meet the challenges of the Run 3 data rates and volumes, the ALICE collaboration is merging the online and offline infrastructures into a common framework: ALICE-$O^2$.
$O^2$ is based on FairRoot and FairMQ, a message-based, multi-threaded and multi-process control framework.
In FairMQ, processes (possibly on different machines) exchange data via message queues either through 0MQ or nanomsg.
In turn, this enables developers to write their reconstruction or analysis process in whatever language they choose or deem appropriate for the task at hand, as long as that programming language can send and receive data through these message queues.
This paper introduces \texttt{fer}, a Go-based toolkit that interoperates with the C++ toolkit FairMQ, to explore the realm of polyglot distributed frameworks.
}
\maketitle
\section{Introduction}
\label{intro}

The ALICE collaboration is merging the online and offline infrastructures into a common framework: ALICE-$O^2$~\cite{ref-aliceo2}.
This work is needed to meet the challenges of the Run 3 data rates and volumes.
$O^2$ is a message-based, multi-threaded and multi-process control framework based on two libraries: FairRoot~\cite{ref-fairroot} and FairMQ~\cite{ref-fairmq}.
Sophisticated processing pipelines can be modeled with FairMQ: router/dealer, request/reply, publish/subscribe, client/server, etc.
The nodes of these pipelines are processes (possibly on different machines) exchanging data via message queues either through \texttt{ZeroMQ}~\cite{ref-zeromq} or \texttt{nanomsg}~\cite{ref-nanomsg}.

From the standpoint of the FairMQ toolkit, the programming language used to implement a given process in the pipeline is irrelevant, as long as data is correctly propagated through the message queues. 
This enables developers to write their reconstruction or analysis process in whatever language they choose or deem appropriate for the task at hand.

This paper presents \texttt{fer}, a Go-based~\cite{ref-golang} library compatible and interoperable with FairMQ.
This paper starts with a brief introduction of the builtin features that make Go a solid choice when dealing with I/O and concurrency.
The principal components of \texttt{fer} and how they interact with C++ FairMQ will then be described.
Finally, Sec.~\ref{sec-fer} will report on the performance (CPU, VMem) of \texttt{fer} and conclude with the main figures of merit of \texttt{fer}, in the context of deployment in a distributed computing setup.

\section{FairMQ concepts}
\label{sec-fairmq-concepts}

\begin{figure}[h]
\centering
\includegraphics[width=0.75\textwidth,clip]{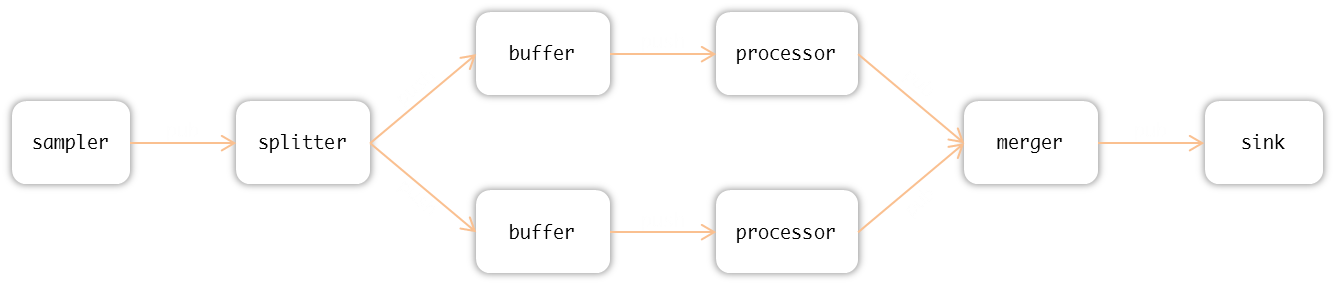}
\caption{Example of a processing pipeline topology, implemented with FairMQ. Each box is a UNIX process, possibly on different machines, connected with other UNIX processes. Data "flows" from left to right.}
\label{fig-topology}
\end{figure}

FairMQ is a distributed processing toolkit, written in \texttt{C++}, with pluggable transports (\texttt{ZeroMQ}, \texttt{nanomsg}, \texttt{shmem}, InfiniBand).
In FairMQ, the bulk of the processing is performed by devices: UNIX processes that are connected with other devices through message queues.
Devices can be connected in various ways, through a wide variety of topologies: router/dealer, request/reply, publish/subscribe, client/server, etc.
The medium used for the connection between devices can be chosen among the following options: \texttt{tcp}, \texttt{udp}, \texttt{ipc}, \texttt{inproc}, shared-memory.
Figure~\ref{fig-topology} shows a typical FairMQ processing pipeline, fanning out data read off a source (\texttt{sampler}) to a pair of processors and then fanning in the data to a single output sink.

FairMQ applications are currently configured via a \texttt{JSON} configuration file, declaring the devices, the ports they are listening on or sending data to, the type and topology of the message queues, etc.
An example of such a configuration file is shown in Fig.~\ref{fig-cfg-json}.

\begin{figure}[h]
\centering
\begin{minted}{json}
{
    "fairMQOptions":
    {
        "devices":
        [{
            "id": "sampler1",
            "channels":
			[{
                "name": "data1",
                "sockets":
				[{
                    "type": "push",
                    "method": "bind",
                    "address": "tcp://*:5555",
                    "sndBufSize": 1000,
                    "rcvBufSize": 1000,
                    "rateLogging": 0
                }]
            }]
        },
[...]
}
\end{minted}

\caption{Example of a JSON configuration file for a FairMQ processing pipeline.}
\label{fig-cfg-json}       
\end{figure}

Topologies in FairMQ can be described in JSON, like in Fig.~\ref{fig-cfg-json}, but other formats and mechanisms are available: XML or via the Dynamic Deployment System (DDS)~\cite{ref-dds}.
Devices are completely agnostic with regard to the technology used to modify, deploy, and push the final configuration.
Once the user has written the correct configuration, declaring inputs and outputs of the whole pipeline, the \texttt{C++} device is launched, is given that configuration and essentially runs an infinite \texttt{for} loop, waiting for data to be received, sending refined data down the pipeline, or waiting for external commands to control the finite state machine of a FairMQ application.
An abridged declaration of the \texttt{C++} base class for devices is shown in Fig.~\ref{fig-fairmq-cxx-device}.

\begin{figure}[h]
\centering
\begin{minted}{c++}
class FairMQDevice : <...> {
public:
 int
 Send(const std::unique_ptr<FairMQMessage>& msg,
      const std::string& chan, const int i) const;

 int
 Receive(const std::unique_ptr<FairMQMessage>& msg,
         const std::string& chan, const int i) const;

protected:
  virtual void Init();    virtual void InitTask();
  virtual void Run();
  virtual bool ConditionalRun();
  virtual void Pause();
  virtual void Reset();   virtual void ResetTask();
};
\end{minted}
\caption{Excerpt of the \texttt{FairMQDevice} class, the main API for implementing a device.}
\label{fig-fairmq-cxx-device}
\end{figure}

Users are supposed to at least override the \texttt{FairMQDevice::Run()} method in their derived classes.
This method is where data is sent to the downstream devices (using the \texttt{FairMQDevice::Send(\ldots)} method) or received from upstream devices (using the \texttt{FairMQDevice::Receive(\ldots)} method).
As a device is usually instantiated within its own UNIX process, users are usually free to leverage any kind of library or programming paradigm inside the boundaries of their own memory address space.

With FairMQ's microservice-like architecture, processing pipelines can horizontally scale much more easily when data taking or data processing demand it.
Indeed, it is just a matter of adding more processing resources to the pool and connect them via message queues to accomodate a change in the processing load.
Abstracting away the minute details of how devices are connected and distributed (all in the same address space, on different machines, etc.) allows users to very easily isolate a faulty device and mitigate its impact on others, \emph{e.g.} preventing it from corrupting memory of other devices.
This also allows to use processes' boundaries to segregate components that are multithread friendly from others that are not.

On one hand, requiring that data is communicated between devices through these message queues healthly constrains the set of \texttt{C++} features one can use to implement the data model.
Indeed, the data being exchanged will tend to resemble plain old data (POD).
On the other hand, relying on off the shelf libraries such as \texttt{ZeroMQ} or \texttt{nanomsg} to implement the actual data transport between devices, enables the implementation of devices in any language that has support for \texttt{ZeroMQ} or \texttt{nanomsg}.
A FairMQ pipeline could theoretically be composed of $n$ devices, each device implemented in a different programming language, presumably one that fits best the task at hand.

In the following, the feasability of a \texttt{Go}~\cite{ref-golang} based toolkit, interoperable with FairMQ, \texttt{fer}, is investigated.

\section{Elements of Go}
\label{sec-intro-go}
Go is an open source general programming language with built-in support for concurrency programming.
It has been released in 2009 under the BSD-3 license.
Go has a static type system with first-class functions, closures, and interfaces.
The general syntax of Go is close to that of C and C++.
Even if Go programs are statically typed, code verbosity is limited thanks to Go's type inference system.
Go's concurrency primitives derive from Hoare's Communicating Sequential Processes (CSP)~\cite{ref-csp}: \emph{goroutines} and \emph{channels}.
Channels are typed conduits that connect goroutines together.
Goroutines are very lightweight green threads: it is possible to schedule thousands of them on regular hardware.
This is achieved thanks to the Go runtime multiplexing goroutines on OS threads and the ability to grow and shrink each goroutine's stack as needed.
The last element for easy concurrency programming is the garbage collector that takes care of freeing memory from the correct thread, when it is not needed anymore.

Go is available on all major platforms (Linux, Windows, macOS, Android, iOS, etc.) and for many architectures (\texttt{amd64}, \texttt{arm64}, \texttt{i386}, \texttt{s390x}, \texttt{mips64}, etc.).
The Go toolchain can very easily cross-compile from one OS/Arch pair to any other pair, by just modifying two environment variables.
The Go toolchain generates completely statically compiled binaries.
This enables a very efficient deployment model where binaries produced on a developer laptop can be quickly provisioned on a bare metal production cluster.
Finally, compiling Go code is not only fast thanks to its package and import system, it is also regular and automatically handles dependencies, recursively discovering and installing them.
Indeed, installing Go packages is usually just a matter of entering in the shell:

\begin{minted}{bash}
	%> go get -v github.com/sbinet-alice/fer
	github.com/sbinet-alice/fer/mq
	github.com/pkg/errors
	golang.org/x/net/context
	github.com/sbinet-alice/fer/config
	nanomsg.org/go-mangos
	[...]
	nanomsg.org/go-mangos/transport/tcp
	github.com/sbinet-alice/fer/mq/nanomsg
	github.com/sbinet-alice/fer/mq/zeromq
\end{minted}

A single command that works on all platforms and operating systems supported by the Go toolchain.

\section{Design and components of \texttt{fer}}
\label{sec-fer}

As introduced in Sec.~\ref{sec-fairmq-concepts}, implementing a toolkit compatible with FairMQ requires to be able to configure FairMQ-like devices, using \emph{e.g.} the JSON configuration file format; to create topologies of devices according to the configuration and connected via a transport library (\emph{e.g.} \texttt{ZeroMQ} or \texttt{nanomsg}); and finally to execute the main routines of the devices.

The requirements above are met by \texttt{fer}~\cite{ref-fer}, a toolkit implemented in Go.
Similar to its \texttt{C++} sibling, \texttt{fer} exposes the concept of a device by way of the \texttt{Device} interface, as shown in Fig.~\ref{fig-fer-device}.

\begin{figure}[h]
\centering
\begin{minted}{go}
package fer

import "github.com/sbinet-alice/fer/config"

// Device is a handle to what users get to run via the Fer toolkit.
type Device interface {
	Run(ctl Controler) error
}

type DevConfigurer interface { Configure(cfg config.Device) error }
type DevIniter     interface { Init(ctl Controler) error }
type DevPauser     interface { Pause(ctl Controler) error }
type DevReseter    interface { Reset(ctl Controler) error }
\end{minted}
	\caption{Excerpt of the \texttt{fer} package, with the main interfaces it exposes. Users are only required to implement the \texttt{Device} interface, other interfaces are optional.}
\label{fig-fer-device}
\end{figure}

A type implements the \texttt{Device} interface as soon as it has a method named \texttt{Run()} that takes a \texttt{Controler} interface and returns an \texttt{error}.
The \texttt{Controler} interface is shown in Fig.~\ref{fig-fer-controler}.
It is used to retrieve named channels bound to incoming or outgoing data, through the \texttt{Chan()} method.
An additional \texttt{Done()} method exposes transitions in the \texttt{fer} final state machine (\texttt{RUN}, \texttt{PAUSE}, \texttt{STOP}, etc.)

\begin{figure}[h]
	\centering
	\begin{minted}{go}
// Controler controls devices execution and gives a device access to input and
// output data channels.
type Controler interface {
	Logger
	Chan(name string, i int) (chan Msg, error)
	Done() chan Cmd
}

// Msg is a quantum of data being exchanged between devices.
type Msg struct {
	Data []byte // Data is the message payload.
	Err  error  // Err indicates whether an error occured.
}

// Cmd describes commands to be sent to a device, via a channel.
type Cmd byte
	\end{minted}
	\caption{Declaration of the \texttt{Logger} and \texttt{Controler} interfaces in the \texttt{fer} package.
	\texttt{Controler}s expose communication channels that can exchange data messages, \texttt{Msg}, that embed the actual payload or an error if any.
}
	\label{fig-fer-controler}
\end{figure}

The input configuration is parsed by the \texttt{fer/config} package and passed to the devices via the \texttt{DevConfigurer} interface, if that device implements it.

\subsection{Implementing a \texttt{fer} device}
This section shows how implementing a \texttt{fer} device and integrating it inside a \texttt{C++} pipeline can be done.
Figure~\ref{fig-fer-processor} shows how one can implement the \texttt{processor} from Fig.~\ref{fig-topology}.
The \texttt{processor} retrieves two channels, the \texttt{"data1"} input channel and the \texttt{"data2"} output channel.
The \texttt{Run()} method contains an infinite \texttt{for} loop.
At each iteration, the device waits for data coming in from either the \texttt{idatac} or the \texttt{Done()} channel.
When the input channel is ready -- a message has been delivered to the device -- the associated payload is modified by the device (just appending some data) and then sent downstream through the output channel \texttt{odatac}.
Concurrently, if any command is sent on the \texttt{Done()} channel, the \texttt{for} loop exits and the Go runtime can reclaim resources used by the associated goroutine.

\begin{figure}[h]
	\centering
\begin{minted}{go}
package mydevice

import (
	"github.com/sbinet-alice/fer"
	"github.com/sbinet-alice/fer/config"
)

type processor struct {
	cfg    config.Device
	idatac chan fer.Msg
	odatac chan fer.Msg
}

func (dev *processor) Configure(cfg config.Device) error {
	dev.cfg = cfg
	return nil
}

func (dev *processor) Init(ctl fer.Controler) error {
	idatac, err := ctl.Chan("data1", 0) // handle err
	odatac, err := ctl.Chan("data2", 0) // handle err
	dev.idatac = idatac
	dev.odatac = odatac
	return nil
}

func (dev *processor) Run(ctl fer.Controler) error {
	str := " (modified by "+dev.cfg.Name()+")"
	for {
		select {
		case data := <-dev.idatac:
			out := append([]byte(nil), data.Data...)
			out = append(out, []byte(str)...)
			dev.odatac <- fer.Msg{Data: out}
		case <-ctl.Done():
			return nil
		}
	}
}
\end{minted}
\caption{Implementation of the \texttt{DevConfigurer}, \texttt{DevIniter} and \texttt{Device} interfaces.}
\label{fig-fer-processor}
\end{figure}

Once a type implements the \texttt{fer.Device} interface, a value of that type can be created and passed to the \texttt{fer.Main} function that takes care of the minute details of scheduling devices, establishing socket connections and forwarding external commands.
This simplifies the creation of the main entry point for the whole program, as shown in Fig.~\ref{fig-fer-main}.

\begin{figure}[h]
	\centering
\begin{minted}{go}
package main

func main() {
	err := fer.Main(&processor{})
	if err != nil {
		log.Fatal(err)
	}
}
\end{minted}
	\caption{Main program for the \texttt{processor} example device. For brevity, imports are omitted.}
	\label{fig-fer-main}
\end{figure}

Compiling and running this processor is done via:
\begin{minted}{bash}
%> go get ./my-device
%> $GOPATH/bin/my-device --id processor --mq-config ./path-to/config.json
\end{minted}

\subsection{Go implementation of \texttt{ZeroMQ}}

At the beginning of the development of \texttt{fer}, a pure Go implementation of the \texttt{nanomsg} package was readily available but not for \texttt{ZeroMQ}.
Only a so-called \texttt{cgo} based package existed, one that linked to the C/C++ \texttt{ZeroMQ} library and wrapped the C/C++ calls with Go functions and types.
Calling Go functions that call C/C++ and back is not only costly -- around ten times a normal function call -- it also breaks the easy compilation, installation, and deployment promises that Go users are accustomed to.
To correct this deficiency, a pure Go package that implements the required subset of the \texttt{ZeroMQ} protocol and sockets to provide interoperability with the C++ implementation was developed: \texttt{go-zeromq/zmq4}~\cite{ref-go-zmq}.

\begin{figure}[h]
\centering
\includegraphics[width=\textwidth,clip]{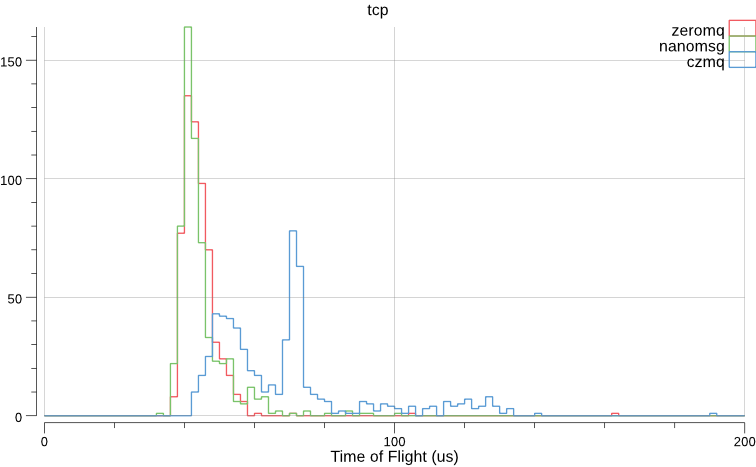}
	\caption{Distribution of the time of flight of a single token (\texttt{uint64}) through a pipeline \texttt{p = \{ source|processor|sink\}}, for different transports and protocols: \texttt{zeromq}, \texttt{nanomsg} and \texttt{czmq}.
	\texttt{czmq} is the C/C++ \texttt{ZeroMQ} library called from Go.}
\label{fig-tof-tcp}
\end{figure}

As shown in Fig.~\ref{fig-tof-tcp}, the mean value for the time of flight of a single \texttt{uint64} token exchanged between a source device, a processor device and a sink device, goes from $\sim 70~\mu$s for the \texttt{czmq} C++ version (called from Go) down to $\sim 45~\mu$s for the pure Go \texttt{nanomsg} and \texttt{zeromq} transports.
The maximum resident set size (\texttt{MaxRSS} from \texttt{/proc/self/statm}) for all configurations was around 22~Mb.

\section{Conclusions}
\label{sec-conclusions}

The ALICE-$O^2$ framework based on the message passing infrastructure from FairMQ enables one to build applications that resemble microservices architectures.
These types of architectures allow systems to scale horizontally quite easily, by just adding new processor devices and fanning out data to these new processors, in \emph{e.g.} a round-robin fashion.
FairMQ, and microservices in general, is language agnostic: developers can thus implement their device in whatever language they deem adequate for the task at hand.
This paper introduced \texttt{fer}, a Go-based toolkit that interoperates with the C++ toolkit FairMQ, to explore the realm of polyglot distributed frameworks.
The \texttt{fer} toolkit provides a fast, "one-command" installation procedure that works on all platforms and operating systems supported by the Go toolchain.
It also provides Grid-friendly deployments thanks to its single, statically compiled, binary.
As such it can be a great testbed environment to explore new avenues.

\end{document}